\documentclass[12pt]{article}

\usepackage{epsf}
\usepackage{amsmath,amssymb}
\usepackage{latexsym}
\usepackage{graphicx,color}
\usepackage{wrapfig}
\usepackage{latexsym}
\usepackage{graphics}

\usepackage{color}
\usepackage{delarray,color}
\usepackage{fancybox}  

\newcommand {\beq}{\begin{eqnarray}}
\newcommand {\eeq}{\end{eqnarray}}

\newcommand{\be}{\begin{equation}}
\newcommand{\ba}{\begin{eqnarray}}
\newcommand{\ea}{\end{eqnarray}}
\newcommand{\ee}{\end{equation}}

\newcommand{\beqa}{\begin{eqnarray}}
\newcommand{\eeqa}{\end{eqnarray}}
\newcommand{\CR}{\nonumber \\}

\newcommand{\unit}{\hbox to 3.8pt{\hskip1.3pt \vrule height 7.4pt
    width .4pt \hskip.7pt \vrule height 7.85pt width .4pt \kern-2.4pt
    \hrulefill \kern-3pt \raise 3.7pt\hbox{\char'40}}}

\def\matt[#1,#2,#3,#4]{\left(%
\begin{array}{cc} #1 & #2 \\ #3 & #4 \end{array} \right)}

\usepackage{tabularx}

\catcode`\@=11
\@addtoreset{equation}{section}

\catcode`@=12
\relax

\newcommand{\tr}{\mathrm{Tr}}

\begin{document}

\begin{titlepage}

\setcounter{page}{0}

\renewcommand{\thefootnote}{\fnsymbol{footnote}}

\begin{flushright}
YITP-14-79 \\
\end{flushright}

\vskip 1.35cm

\begin{center}
{\Large \bf 
A Localization Computation 
in Confining Phase
}

\vskip 1.2cm 

{\normalsize
Seiji Terashima\footnote{terasima(at)yukawa.kyoto-u.ac.jp}
}

\vskip 0.8cm

{ \it
Yukawa Institute for Theoretical Physics, Kyoto University, Kyoto 606-8502, Japan
}

\end{center}

\vspace{12mm}

\centerline{{\bf Abstract}}

In this note we show that 
the gaugino condensation of 4d ${\cal N}=1$ supersymmetric 
gauge theories in the confining phase can be computed
by the localization technique with 
an appropriate choice of a supersymmetry generator.

\end{titlepage}
\newpage

\tableofcontents
\vskip 1.2cm 

\section{Introduction and Summary}

An analytic computations in quantum field theories
are usually hopeless except in some special classes of them.
The supersymmetric (SUSY) gauge field theories 
are in the special class.
The early example of such computations in the SUSY field theories
is the Witten index \cite{Witten1}.
Recently, the localization computations in the SUSY field theories 
on curved space initiated by Pestun \cite{Pestun}
have been investigated intensively, 
which can be considered as generalizations of
the topological field theories \cite{Witten2} and
the Nekrasov partition function \cite{Nekrasov}.
The localization computations use deformations 
of the action which do not change the partition function 
and some correlation functions 
and have been generalized to other 4d manifolds \cite{HH}-\cite{Imamura} and 
also SUSY theories in various
dimensions (see, for examples, \cite{Kap}-\cite{Sugishita}).

The other powerful technique to compute exactly some correlators
in SUSY gauge theories is the holomorphy of the ``superpotential''
(see the review \cite{IS1}).
Indeed, Seiberg 
found the exact low energy superpotential 
of 4d ${\cal N}=1$ SUSY $SU(N_c)$ QCD on ${\bold R^4}$ \cite{Seiberg1}
by using 
the holomorphy of the superpotential written in the superfields 
and the result of \cite{ADS}.
In particular, the gaugino condensation, i.e. 
the v.e.v of the gaugino bi-linear,
was exactly computed for the SUSY QCD with massive chiral multiplets.
It was also computed in \cite{NSVZ85}.
This technique was applied to the prepotential of 
4d ${\cal N}=2$ SUSY gauge theories 
by Seiberg and Witten \cite{SW}.

One might expect that the localization technique is powerful enough 
to compute any exactly computable quantity in SUSY gauge theories.
For example, the Seiberg-Witten prepotential can be 
reproduced from the Nekrasov partition function.
However,
the above mentioned gaugino condensation is an obvious exception
for this expectation so far.
Indeed, the localization technique has not been applied to 
a SUSY gauge theory in the confining phase nor
a computation of a v.e.v. of local operators like the gaugino bi-linear.
Thus, it would be important and interesting to 
compute the gaugino condensation in 
4d ${\cal N}=1$ SUSY $SU(N_c)$ (massive) QCD
by the localization technique directly
because this will open up other interesting 
applications of the technique.

In this paper, we will show that 
the gaugino condensation can actually be computed
by the localization technique with 
an appropriate choice of a SUSY generator.
We consider the theory on $S^1 \times {\bold R}^3$
and then take the limit to ${\bold R}^4$
.\footnote{
Another simplest one is 4d ${\cal N}=1$ SUSY gauge theory on $S^4$
which we construct explicitly in the appendix.
The computation of gaugino condensation of this theory
would be related to the strong coupling instanton \cite{Amati}
because the instanton in ${\bold R}^4$ is related to the one in $S^4$.
However, the localization technique can not be applied to this theory,
at least naively,
as we can see in the appendix. 
We will also argue that the direct application of 
the localization
technique to the theory on ${\bold R^4}$
can not give the weak coupling limit.}
The actual computation for this theory is almost identical 
to the one performed by Davies et. al. \cite{Davies, DHK},
in which they consider the small radius limit of $S^1$
and argued that the gaugino condensation in the ${\cal N}=1$ SUSY
Yang-Mills theory is independent of the radius of $S^1$.
Furthermore, it was shown in \cite{DHK}
that the value of the gaugino condensation agrees 
with the one obtained in \cite {ADS, Seiberg1} 
using the holomorphy.
Therefore, we reproduce the gaugino condensation
using the localization technique,\footnote{
Precisely speaking, the localization computation 
can be done for $S^1 \times {\bold R}^3$.
However, the gaugino condensation of the theory on ${\bold R^4}$ can be
obtained by the large radius limit.
Note that the theory on $S^1 \times {\bold R}^3$
is three dimensional theory in the IR limit, however,
the scale is very small 
for the large radius.
Thus, at a scale between this scale and the dynamical scale $\Lambda$
the theory is in a 4d confining phase
and the radius of $S^1$ cab be regarded as an infra-red regulator.
} 
and we hope that the discussions in the paper
can be generalized to broader classes of SUSY gauge theories
with only minor changes.
Indeed, 
we believe that we can apply the arguments in this paper to
generic ${\cal N}=1$ SUSY gauge theories 
and compute the chiral v.e.v. in the weak coupling limit.
This is why we think the results in this paper are important
although the explicit computations in the paper 
was essentially known as stated above.

The organization of this paper is as follows:
In section 2 we first briefly review the 
4d ${\cal N}=1$ SUSY gauge theory on $S^1 \times S^3$.
Then, we show that by the localization technique
the correlators of the lowest components 
of the chiral multiplets can be computed 
semi-classically around the anti-self-dual configurations.
In section 3 we compute the gaugino condensation
for the pure SUSY Yang-Mills on $S^1 \times {\bold R}^3$
using the localization technique.
We comment on the inclusion of the chiral multiplets shortly
in section 4.
In the appendix, we construct the 
4d ${\cal N}=1$ SUSY gauge theory on $S^4$
although the localization technique is not useful for it.

\section{4d ${\cal N}=1$ SUSY Gauge Theory on $S^1 \times S^3$ 
and Localization}

In this section, 
we will briefly review 
the 4d ${\cal N}=1$ SUSY gauge theory on $S^1 \times S^3$ 
\cite{Imamura,Dolan, Gadde, Martelli}\footnote{We will closely follow the notations
and conventions in \cite{Hosomichi}.}
and then show that 
the path-integral for the correlation function of the 
lowest components of chiral multiplets 
can be reduced 
to the semi-classical computations around the instantons
by adding an appropriate regulator action.
Of course, the discussion in this section 
can be applied to the 4d ${\cal N}=1$ SUSY gauge theory on $S^1 \times {\bold R}^3$
which is our main concern. 
Note that the SUSY theory with gaugino condensation can
not be put on $S^1 \times S^3$ without spontaneous breaking of the SUSY
because of the required continuous R-symmetry.\footnote{
We thank an editor of a journal for pointing out this point.}
However, there would be some ${\cal N}=1$ 
SUSY gauge theories on $S^1 \times S^3$ which 
can have computable non trivial v.e.v. of the chiral multiplets
by the localization technique.
Thus, we will present the discussion in this section 
for the SUSY gauge theory on $S^1 \times S^3$.

First, let us remind that 
the SUSY transformation of the vector multiplet 
for Euclidian 4d ${\cal N}=1$ SUSY gauge theory on ${\bold R}^4$ is
\begin{eqnarray}
 \delta A_m &=& \frac{i}{2} (\epsilon \sigma_m \bar{\lambda} -
\bar{\epsilon} \bar{\sigma_m} \lambda), \nonumber \\
\delta \lambda &=& \frac{1}{2} \sigma^{mn} \epsilon F_{mn} - \epsilon D, \nonumber \\
\delta \bar{\lambda} &=& \frac{1}{2} \bar{\sigma}^{mn} \bar{\epsilon} F_{mn} 
- \bar{\epsilon} D, \nonumber \\
\delta D &=& - \frac{i}{2} \epsilon \sigma^m D_m \bar{\lambda} 
-\frac{i}{2} \bar{\epsilon} \bar{\sigma}^m D_m \lambda,
\label{susytrf}
\end{eqnarray}
where $m=1,2,3,4$, all fields are in the adjoint representation of the gauge group
$G$ and $\epsilon, \bar{\epsilon}$ are constant spinors.
We have introduced the chiral decomposed Gamma matrix $\sigma^m$:
$\sigma^a$ is the Pauli matrix for $a=1,2,3$ and 
$\sigma^4=i$, and $\bar{\sigma}^m=(\sigma^m)^\dagger$.
We also defined 
$\sigma_{mn}=\frac{1}{2} (\sigma_m \bar{\sigma}_n -\sigma_n
\bar{\sigma}_m) $
and 
$\bar{\sigma}_{mn}=
\frac{1}{2} (\bar{\sigma}_m \sigma_n- \bar{\sigma}_n \sigma_m)$.
The spinor indices are raised or lowered by $\epsilon^{\alpha \beta}$
which has $\epsilon^{12}=-\epsilon_{12}=1$.\footnote{
The normalizations of the $\sigma_{mn}, \bar{\sigma}_{mn}$ 
and the fermions are different from the one in the Wess-Bagger's text.}

It is known \cite{Imamura, Hosomichi} that the SUSY transformation (\ref{susytrf}) is consistent 
with the 4d ${\cal N}=1$ SUSY gauge theory on 
$S^1 \times S^3$ if the derivative $D_m$
is defined as follows:
\begin{eqnarray}
 D_m = 
{\cal D}_m -i q V_m
,
\label{dd}
\end{eqnarray}
where ${\cal D}_m$ is the covariant derivative
including the 
spin connection, for example 
${\cal D}_m = \partial_m +\frac{1}{4}w_m^{ab} \gamma^{ab}$
for spinors.
Here  the $V_m$ is an appropriate background field and
the $q$ is the R-charge for which
we assigned $q(\epsilon)=1, q(\bar{\epsilon})=-1,q(\lambda)=1,  
q(\bar{\lambda})=-1$.
For simplicity, we will consider the unit 3-sphere and 
$ds^2 = dt^2+ds^2_{S^3}$
where $0 \leq t < 2 \pi R$ is the periodic coordinate for $S^1$.
Then, the background field is fixed to be
\begin{eqnarray}
 V_m dx^m=-\frac{i}{2} dt,
\end{eqnarray}
and the Killing spinors are given by the solutions of the following equations:
\begin{eqnarray}
 D_m \epsilon = - \frac{1}{2} \sigma_m \bar{\sigma}_4 \epsilon, \,\,\,\,
 D_m \bar{\epsilon} = - \frac{1}{2} \bar{\sigma}_m \sigma_4  \bar{\epsilon}.
\end{eqnarray}


It should be stressed that
we regard 
$\lambda$ and $\bar{\lambda}$ are independent holomorphic 
fermionic 2-components spinor fields in the 4d Euclidian spacetime.
Indeed, these two spinors are fundamental representations of 
the former and the latter $SU(2)$ of
$SU(2) \times SU(2)(=Spin(4))$, thus they can not be related by the
complex conjugation.
Because any reality condition can not be imposed on the fundamental
representation of $SU(2)$, we regard $\lambda, \bar{\lambda}$ as 
the formal holomorphic path-integral variables.
The SUSY parameters $\epsilon$ and $\bar{\epsilon}$ 
are also independent fields.

Usual ${\cal N}=1$ SUSY Yang-Mills action with the theta term
\begin{eqnarray}
 L_g= {\rm Tr} \left[ 
\frac{1}{ g^2} \left( \frac{1}{2} F_{mn} F^{mn} +D^2 +i \bar{\lambda}
		     \bar{\sigma}^m D_m \lambda \right) +i
 \frac{\theta}{16 \pi^2} F \tilde{F} \right],
\label{sym}
\end{eqnarray}
is SUSY invariant on $S^1 \times S^3$ with the covariant derivative 
$D_m$ defined by (\ref{dd}).
We will also use the complexified coupling constant
$\tau \equiv  \frac{\theta}{2 \pi}+\frac{4 \pi i}{g^2}$.

Now we are applying the localization technique
to the theory.
In this paper, the localization technique simply means the
use of the following identity
for the operators satisfying $\delta {\cal O}_a=0$:
\begin{eqnarray}
\frac{d}{dt} \langle {\cal O}_1 {\cal O}_2 \cdots {\cal O}_n
e^{-t \delta \int V} \rangle =0,
\label{e1}
\end{eqnarray}
where $\delta$ is a symmetry transformation of the theory
and $V$ should satisfies
$\delta^2 \int_{S^3 \times S^1} V=0$.
With this identity we can compute the correlator 
$\langle {\cal O}_1 {\cal O}_2 \cdots {\cal O}_n \rangle$
with a sufficiently large $t$ 
if the real part of $\delta \int_{S^3 \times S^1} V$ is non-negative.

In order to use the localization technique,
we should choose a SUSY transformation $\delta$ and a regulator
Lagrangian $V$.
Here the choice of $\delta$ means the choice of $\epsilon$ and
$\bar{\epsilon}$. 
As in \cite{Pestun}, 
we simply take the regulator Lagrangian as following:
\begin{eqnarray}
 V= (\delta \lambda)^\dagger \lambda +
(\delta \bar{\lambda})^\dagger \bar{\lambda},
\label{defV}
\end{eqnarray} 
where we should define $(\delta \lambda)^\dagger$ 
appropriately.

First, let us assume both $\epsilon$ and $\bar{\epsilon}$ are nonzero.
Then, the bosonic contributions from the first term is
\begin{eqnarray}
 (\delta \lambda)^\dagger \delta \lambda  
\sim F^{+}_{mn} F^{+mn} +D^2,
\end{eqnarray}
where we have used
\begin{eqnarray}
 \delta \lambda &=& 
\frac{1}{2} \sigma^{mn}  \epsilon F^{+}_{mn} - \epsilon  D,
\end{eqnarray}
and $F^{+} \, (F^-)$ are the anti-self-dual (self-dual) part of 
the field strength $F$,
respectively.
Then, the saddle points for $ t \rightarrow \infty$
will satisfy $F^+=0$ and $D=0$.
Because the contribution from the other term in $V$, i.e.
$(\delta \bar{\lambda})^\dagger \delta \bar{\lambda}  
\sim (F^{-})_{mn} (F^{-})^{mn} +D^2$,
will give $F^-=0$,
we conclude that 
the saddle points satisfy $F_{mn}=0$ and $D=0$
where the regulator action is essentially the Yang-Mills action.
Thus the partition function can be calculated 
in the weak coupling limit.
This is well-known result.
Indeed, with the twist of the boundary condition along the $S^1$,
the partition function on $S^1 \times S^3$ is 
the superconformal index \cite{Romels, Maldacena, Dolan, Gadde, Imamura}.

In this paper, we consider another possibility for the localization 
computation on $S^1 \times S^3$:
We take $\bar{\epsilon} \neq 0$, but $\epsilon=0$, 
which implies $\delta \delta=0$. 
Thus the condition $\delta \delta \int V=0$ is 
trivially satisfied.\footnote{
However, this means that 
we can not use the equivariant index theorem
for the 1-loop computation 
even though the path-integral will be reduce to the 
semi-classical instanton calculation.
The instantons do not localize to somewhere, for example,
north or south poles.
}
Explicitly, the SUSY transformation is
\begin{eqnarray}
 \delta A_m &=& -\frac{i}{2} \bar{\epsilon} \bar{\sigma_m} \lambda, \nonumber \\
\delta \lambda &=& 0, \,\,\,\,\,\,
\delta \bar{\lambda} = \frac{1}{2} \bar{\sigma}^{mn} \bar{\epsilon} F_{mn} 
- \bar{\epsilon} D, \nonumber \\
\delta D &=& -\frac{i}{2} \bar{\epsilon} \bar{\sigma}^m D_m \lambda,
\label{susytrf2}
\end{eqnarray}
With this $\delta$, we have $\delta \lambda=0$, thus 
a correlation function of any gauge invariant 
combinations of $\lambda$ can be computed by the localization technique.
The Killing spinor can be taken explicitly as 
\begin{eqnarray}
 \bar{\epsilon}=\frac{1}{\sqrt{2}}
\left( 
\begin{array}{c} e^{\frac{i}{2} (-\chi+\phi+\theta)} \\ 
e^{\frac{i}{2} (-\chi+\phi-\theta)} 
\end{array} 
\right),
\end{eqnarray}
where we use the coordinate system with the metric $ds^2=\cos^2
\theta d \phi^2+\sin^2 \theta d \chi^2 +d \theta^2$.
Taking $V$ as in (\ref{defV}),
we easily see that 
\begin{eqnarray}
 V = F^{-}_{mn} F^{-mn} 
+D^2=\frac{1}{2} F_{mn} F^{mn} 
+\frac{1}{2} F_{mn} \tilde{F}^{mn}
+D^2,
\label{regve}
\end{eqnarray}
where $\tilde{F}$ is the dual of the field strength, and
the saddle points equations are $F^-=0$ and $D=0$, i.e.
the instantons (anti-self-dual connections) on $S^1 \times S^3$.\footnote{
This saddle point equations was also noted in a recent paper
\cite{Martelli}.}
Note that this includes the Yang-Mills action with a pure imaginary $\theta$.
This also means that by taking $t \rightarrow \infty$ limit
the theory is arbitrary weak coupling because 
only the Yang-Mills action affects the fluctuations around the instantons.
The value of the Lagrangian (\ref{sym}) at the saddle points 
is $ 2 \pi i \tau \times (\mbox{instanton number})$ 
because $F=\tilde{F}$.
It is interesting that there is a coupling constant dependence
in the computation using the localization for the ${\cal N}=1$ SUSY 
gauge theory on $S^1 \times S^3$.

For the chiral multiplets on $S^1 \times S^3$,
the SUSY transformation were given as
\begin{eqnarray}
 \delta \phi &=& - \epsilon \psi, \,\,\,\,\,
\delta \psi =i \sigma^m \bar{\epsilon} D_m \phi 
+\frac{3 ir }{4} \sigma^m D_m \bar{\epsilon} \phi + \epsilon F, 
\nonumber \\
 \delta \bar{\phi} &=&  +\bar{\epsilon} \bar{\psi}, \,\,\,\,\,
\delta \bar{\psi} =i \bar{\sigma}^m \epsilon D_m \bar{\phi} 
+\frac{3 ir }{4} \bar{\sigma}^m D_m \epsilon \bar{\phi} + \bar{\epsilon} \bar{F}, 
\nonumber \\
 \delta F &=& +i \bar{\epsilon} \bar{\sigma}^m D_m \psi 
+\frac{i(3r-2) }{4}  D_m \bar{\epsilon} \bar{\sigma}^m \psi 
-i \bar{\epsilon} \bar{\lambda} \psi, 
\nonumber \\
 \delta \bar{F} &=& -i \epsilon \sigma^m D_m \bar{\psi} 
-\frac{i(3r-2) }{4}  D_m \epsilon \sigma^m \bar{\psi} 
-i \epsilon \bar{\psi} \lambda.
\end{eqnarray}
where $r$ is the R-charge of $\phi$.
Under the SUSY transformation,
the following kinetic term is invariant:
\begin{eqnarray}
 {\cal L}_m &=& D_m \bar{\phi} D^m \phi+\frac{1}{2} (3r-2)
(D_4 \bar{\phi} \phi -\bar{\phi} D_4 \phi)
+\frac{3 r(-3 r +4)}{4}  \bar{\phi} \phi
-i \bar{\phi} D \phi \nonumber \\
&& -i \bar{\psi} \bar{\sigma}^m D_m \psi
-i \frac{3r-2}{2} \bar{\psi} \bar{\sigma}_4 \psi 
+ i \bar{\psi} \bar{\lambda} \phi + i \bar{\phi} \lambda \psi +\bar{F}F.
\end{eqnarray}
We can easily see that the usual superpotential terms are the SUSY invariant
if the R-charge is $2$.

For $\epsilon=0$, 
we have $\delta \phi=0, \delta \bar{F}=0$.
Thus any gauge invariant combination 
of the lowest components of the chiral multiplets
is invariant under the SUSY transformation generated by the $\delta$
with $\epsilon=0$.
Now let us consider the following $\delta$-exact Lagrangian:
\begin{eqnarray}
 L_{\bar{W}}= \delta \left( 
\frac{\partial \bar{W}(\bar{\phi})}{\partial \bar{\phi}^i} 
\bar{\eta} \bar{\psi}^i 
\right),
\end{eqnarray} 
where $\bar{\eta} $ is 
a Grassmann even spinor 
defined by 
\begin{eqnarray}
\bar{\eta} \equiv U 
 \left( 
\begin{array}{c} 0 \\ 
1
\end{array} 
\right),
\end{eqnarray}
where 
\begin{eqnarray}
 U=\frac{1}{\sqrt2} \left( 
\begin{array}{cc} e^{\frac{i}{2} (-\chi+\phi+\theta)}  & 
-e^{-\frac{i}{2} (-\chi+\phi-\theta)} \\ 
e^{\frac{i}{2} (-\chi+\phi-\theta)} &
e^{-\frac{i}{2} (-\chi+\phi+\theta)}  
\end{array} 
\right),
\end{eqnarray}
is an $SU(2)$ matrix.
Note that 
$\bar{\epsilon}=U 
 \left( 
\begin{array}{c} 1 \\  0
\end{array} 
\right)$
and then we have
$\bar{\eta} \bar{\epsilon}=1$.
This does not seem to keep the rotational symmetry of $S^3$
because of the presence of $\bar{\eta}$,
however, this keeps it.  
Indeed, we can see that 
this Lagrangian is the anti-chiral part of the superpotential
$W(\phi)$.\footnote{
For the fermionic part, 
we have 
\begin{eqnarray}
 \frac{\partial \bar{W}(\bar{\phi}) }{\partial \bar{\phi}^i  
\partial  \bar{\phi}^j} 
(\bar{\epsilon} \bar{\psi}^j )
(\bar{\eta} \bar{\psi}^i )
=  \frac{1}{2} \frac{\partial \bar{W}(\bar{\phi}) }{\partial \bar{\phi}^i  
\partial  \bar{\phi}^j} 
\left(
(\bar{\epsilon} \bar{\psi}^j )
(\bar{\eta} \bar{\psi}^i )
+
(\bar{\epsilon} \bar{\psi}^i )
(\bar{\eta} \bar{\psi}^j )
\right)
\sim 
\frac{1}{2} \frac{\partial \bar{W}(\bar{\phi}) }{\partial \bar{\phi}^i  
\partial  \bar{\phi}^j} 
(\bar{\psi^j} \bar{\psi^i)},
\label{regch}
\end{eqnarray}
where the last equation can be shown by 
taking $\epsilon^1=1, \epsilon^2=0$.
Thus the $L_{\bar{W}}$ is the anti-chiral part of the superpotential.
}
Thus, the correlators we consider do not depend
on parameters in the anti-chiral superpotential.
This can be regarded as a derivation of the holomorphy
without using the superfields.

Now we consider the following regulator Lagrangian:
\begin{eqnarray}
 V= (\delta \psi)^\dagger \psi +
(\delta \bar{\psi})^\dagger \bar{\psi},
\label{defV}
\end{eqnarray} 
where the Hermite conjugate $(\dagger)$ is defined 
such that it is positive definite.
We have
\begin{eqnarray}
 V \sim |D_m \phi|^2+ r^2 |\phi|^2+|F|^2,
\end{eqnarray}
schematically.
For a field with $r \neq 0$,
the saddle point is trivial, i.e. $\phi=0$ and $F=0$.
However, 
in the flat space limit ($S^1 \times {\bold R}^3 $),
we have 
\begin{eqnarray}
 V \sim |D_m \phi|^2+ |F|^2,
\end{eqnarray}
and the saddle points are $D_m \phi=0$ and $F=0$.
Therefore, in general the v.e.v. of the 
lowest components of the chiral multiplets
can be computed by semi-classical computations
if the theory are in the confining phase with SUSY.\footnote{
Note that the regulator term (\ref{regch}) is a Kaehler potential and 
the (\ref{regve}) is the anti-chiral super potential which
will not affect the v.e.v. of the 
lowest components of the chiral multiplets.}

As we stated at the beginning of this section, 
we need to find a SUSY theory with the non trivial v.e.v. of the chiral
multiplets without breaking the SUSY on $S^1 \times S^3$
in order to apply the discussion in this section.
Furthermore, for an explicit computation of the correlation functions
we need to construct the instantons on $S^1 \times S^3$.
Although 
it is important and interesting to perform these explicitly,
we leave these problems in future.
Instead, we will consider the ${\cal N}=1$ SUSY gauge theory
on $S^1 \times {\bold R}^3$ which is considered as 
a small curvature limit of $S^1 \times S^3 $.

\section{Gaugino Condensation in SUSY Yang-Mills by the Localization}

In this section, 
for simplicity, 
we will consider the ${\cal N}=1$ 
SUSY Yang-Mills theory with gauge group G
which has a simple Lie algebra of rank $r$ 
on $S^1 \times {\bold R}^3$ where the radius of
$S^1$ is $R$. 
Here we require the periodic boundary conditions along $S^1$ for 
all fields. The action is given by (\ref{sym}).
For this theory, the important correlation function which 
can be computed using the localization is the gaugino condensation
\begin{eqnarray}
 \langle \tr \left( \lambda \lambda \right) \rangle,
\end{eqnarray} 
which is known to be non-vanishing 
due to the strong coupling effects, i.e. the confinement effects.
Indeed, this has been computed exactly for the theory on ${\bold R^4}$ 
for the classical groups 
in \cite{NSVZ85, SV, MOS, FP}
which was also computed using the holomorphy of the superpotential.
Furthermore, it was also computed
by using the $R \rightarrow 0$ limit 
for any simple gauge group in \cite{Davies}.
Although $R$ itself is not holomorphic 
variable,
in \cite{Davies}, it was argued that the gaugino condensation 
does not depend on $R |\Lambda|$ because of the holomorphy
for the SUSY Yang-Mills theory and it is only the 
dimensionless combination
the correlator can depend on.

Here, we would like to compute the gaugino condensation
by using the localization technique.\footnote{
For the field theory on a non-compact manifold, 
the weak coupling expansion for 
the $t \rightarrow \infty$ limit could not be valid by 
the IR effects. 
Indeed, if we consider the SUSY Yang-Mills on ${\bold R^4}$ 
and using the localization
technique, the large $t$ will not mean
the weak coupling.
The effective dynamical scale determined by $t$
is very low in the large $t$ limit, however,
the effective coupling of the 4d spectrum below this scale
is not weak coupling.
On the other hand, 
we expect that 
the gauge theory on $S^1 \times {\bold R}^3$
will be weak interacting in the large $t$ limit.
In this case, below the mass scale $1/R$ 
there are 3d fields with non-zero 
v.e.v which breaks the gauge group to the Abelian group,
as we will see later.
Thus, in this case we see that the theory is weak coupling
in the large $t$ limit.
Note that if we take $R \rightarrow \infty$, 
this breaking scale goes to zero.
}
The computation we will see below is essentially same
as the one in \cite{Davies, DHK} and then the followings will be almost
a brief review of \cite{Davies, DHK}.
However, we need to be careful 
to check that the validity of the computation in $R \rightarrow 0$
can be applied to our computation in 
$t \rightarrow \infty$ limit with $R$ fixed finite.
We hope that our localization computation can be generalized 
to more general ones.
In this section, we will use the notation and convention used in \cite{DHK}.

First, we will add the regulator action, whose bosonic part 
is $-t \int ((F^-)^2+D^2)$,
to the original SUSY Yang-Mills action in order to compute 
$ \langle \tr \left( \lambda \lambda \right) \rangle$.
Note that the path-integral measure is defined at $t=0$.
Then, taking $t \rightarrow \infty $ limit,
the path integral localized to the saddle point configurations 
which satisfy $F^-=0$ and $D=0$.
Here, the 1-loop determinant around the saddle points may be trivial 
because we are considering the flat space ${\bold R}^3$
with the Euclidian time.\footnote{
This is not precisely true.
As shown in \cite{Poppitz, Anber}, the 1-loop determinant is non-trivial
because of the asymptotic behaviors of the monopoles, 
however, it only contributes to the K\"ahler potential.
Thus, for the computation of the gauge condensation, 
it can be ignored.}
The saddle points (and every configurations)
are characterized by the Wilson loop along $S^1$, the instanton charge
and the magnetic charges.
The Wilson loop is defined by
\begin{eqnarray}
 \langle \phi \rangle = \lim_{|x^\mu| \rightarrow \infty} 
\int^{2 \pi R}_0 dx_0 A_0, 
\end{eqnarray}
which should not be path-integrated and considered as 
a moduli of the vacua because we consider the theory on $S^1 \times {\bold R}^3$
which is non-compact for the three directions.
Note also that the instanton number $k$ is not necessary integer
because we consider non-compact manifold.
Assuming the generic non-zero Wilson loop,
we have the decomposition of gauge group $G$ to $U(1)^r$
which enable us to define the $r$ different magnetic charges.
Among the saddle point configurations, 
we need the configurations which have precisely 
two zero modes of the fermions $\lambda$ under the anti-self-dual configurations
in order to give a non-zero contribution to 
the gaugino condensation because it is a bi-linear of the fermions.  
These configurations are identified and are called the fundamental monopoles 
in \cite{Davies, DHK}.
These fundamental monopoles 
are usual BPS magnetic monopoles parameterized by $i=1, \cdots, r$ which  
is related to the embedding of $SU(2)$ into $G$, and
the KK monopole of \cite{LeeYi}.
It is known that the 1-instanton without magnetic charges 
on $S^1 \times {\bold R}^3$ can be 
considered as a bound state of these $r+1$ monopoles.

The BPS magnetic monopole 
has magnetic charge $g$ given by $\alpha^*_i$
where $\alpha_i$ is a simple root.
The instanton charge $k$ of it is 
$\frac{1}{2 \pi} \alpha_i^* \cdot \langle \phi \rangle$
and then the classical action is 
\begin{eqnarray}
 S=-i \tau  \alpha^*_i \cdot \langle \phi \rangle.
\end{eqnarray}
Here $\alpha^* \equiv 2 \alpha / (\alpha \cdot \alpha )$.
For the saddle points, the regularized action vanishes by definition, of course.
For the KK monopole,
we have $g=\alpha^*_0$, $k=1+\frac{1}{2 \pi} \alpha_0^* \cdot \langle
\phi \rangle$,
and 
\begin{eqnarray}
  S=-2 \pi i \tau -i \tau  \alpha^*_0 \cdot \langle \phi \rangle,
\end{eqnarray}
where $\alpha_0$ is the lowest root which satisfies 
$\sum_{i=0}^r k_i^* \alpha^*_i=0$ with $k_i^*$ is the Kac labels.

The two fermionic zero modes are given by the SUSY transformation
\begin{eqnarray}
 \lambda_\alpha=\sigma^{mn \beta}_\alpha \xi_\beta F_{mn}
 \rightarrow 4 \pi (S_F \xi)_\alpha \alpha\* \cdot H,
\end{eqnarray}
where the limit means $|x_\mu| \rightarrow \infty$  and 
$S_F=\sigma_\mu x^\mu / (4 \pi |x_\nu|^3)$ is the 
massless fermion propagator in 3d.

Before computing the correlators, we need to find 
the quantum vacua of the theory.
Here we will find the quantum vacua and 
compute the gaugino condensation
at $t \rightarrow \infty$ limit where the theory 
is weak coupling.
Later, we will consider the original theory at $t=0$.

The classical massless fields for the $t \rightarrow \infty$ limit
should be $U(1)^r$ Abelian multiplets with zero KK momentum 
of $S^1$, which can be regarded as 3d fields.
Because in the $t \rightarrow \infty$ limit 
the zero modes and non-zero modes are decoupled each other,
we can forget about the non-zero KK momentum modes.
Thus the bosonic part of them are 
the Wilson loop scalars $\phi$ and dual photon scalars $\sigma$,
which can be combined to $r$ complex scalars
\begin{eqnarray}
 z = i (\tau \phi + \sigma).
\end{eqnarray} 
Thus the $z$ are the classical moduli of vacua.
By checking the SUSY transformations, we can see that 
this $z$ forms the ${\cal N}=1$ chiral multiplets $X$
with a rescaled massless fermions 
$\psi=2^{\frac{5}{2}} \pi^2 \frac{R}{g^2} \lambda$,
where we integrated out the auxiliary fields $D$
and have considered the on-shell multiplets.
The kinetic terms are given by the original action as 
\begin{eqnarray}
 S= \frac{1}{8 \pi R^2}
 \int d^3 x \frac{1}{ {\rm Im} \tau} 
X^\dagger
  X|_{ \theta \theta \bar{\theta} \bar{\theta}},
\end{eqnarray}
where no potential terms appears.

Now we will compute the scalar potential semi-classically
and determine the vacua.
Note that our regularization term is the SUSY Yang-Mills 
term with the non-zero theta term, thus it is 
invariant under all SUSY transformations.
Therefore, instead of the scalar potential, we will compute 
the fermion bi-linear terms which are related to the 
superpotential and the 
scalar potential by the SUSY transformation.
The fermion bi-linear can be non-zero only 
for the fundamental monopole configurations,
on which the original action is evaluated to the following value:
\begin{eqnarray}
 S_j=-2 \pi i \tau \delta_{j0} -\alpha^*_j \cdot \langle z \rangle,
\end{eqnarray}
for the $j$-th fundamental monopole.
Note that the v.e.v. of the dual photon $\sigma$ also contributes
to it because 
the additional action
 $\frac{i}{4 \pi} \int d^3x \epsilon_{\mu \nu \rho} \sigma \partial_\mu
 F_{\nu \rho} $
which forces the Bianchi identity 
gives the boundary term:
\begin{eqnarray}
 \frac{i}{2 \pi} \int_{S^2=\partial ({\bold R}^3)} d x_\mu \sigma B_\mu. 
\end{eqnarray}

The path-integral measure 
of the zero modes of the $j$-th monopole is given by
\begin{eqnarray}
 \int d \mu^{(j)}_{mono} = \frac{2}{(\alpha_j)^2}
\frac{\mu^3 R}{g^2} e^{-S_j} \int d^3 a d \Omega d^2 \xi,
\end{eqnarray}
where $a_\mu$ is the position on ${\bold R}^3$,
$\Omega$ is the $U(1)$ phase and $\xi$ is the Grassmann odd
zero modes.
Note that 
this includes the cut-off scale $\mu$ and
the gauge coupling $g$.
These two are defined by the original theory,
especially $g$ is defined by the original action
because the path-integral measure is defined at 
the original action, i.e. $t=0$.
Using this measure and the asymptotic form of the 
fermionic zero modes,
we have
\begin{eqnarray}
 \langle \lambda_\alpha(x) \otimes \lambda_\beta(0) \rangle
\sim \frac{2^6 \pi^3 \mu^3 R}{g^2 (\alpha_j)^2}
\alpha_j^* \otimes \alpha_j^* e^{-S_j} 
\int d^3 a S_F(x-a)_\alpha^{\,\, \gamma } S_F(a)_{\beta \gamma},
\label{fbl}
\end{eqnarray} 
for $|x| \rightarrow \infty$.
The superpotential which gives the contributions of 
$r+1$ fundamental monopoles is found to be
\begin{eqnarray}
 W(X)=\frac{2 \pi \mu^3 R}{g^2} \left(
\sum_{j=1}^r \frac{2}{(\alpha_j)^2} e^{\alpha_j^* X}
+ \frac{2}{(\alpha_0)^2} e^{2 \pi i \tau +\alpha_0^* X}
   \right).
\end{eqnarray}
The vacua can be fixed by $\frac{d W}{d X}=0$ to
\begin{eqnarray}
 X=\sum_{j=1}^r a_j w_j, \,\,\, \mbox{where} \,\,\,\,
e^{a_j}= \frac{k_j^* (\alpha_j)^2 e^{2 \pi i \tau}}{2 \kappa},
\end{eqnarray}
and
\begin{eqnarray}
 \kappa^{c_2}=e^{2 \pi i (c_2-1) \tau} 
\prod_{j=0}^r \left( \frac{k_j^* (\alpha_j)^2}{2} \right)^{k_j^*},
\end{eqnarray}
which has $c_2$ roots corresponding to $c_2$ vacua.
Here $c_2= \sum_{i=0}^r k_i^*$ is the dual Coxeter number.

We can compute the gaugino condensation
by evaluating the (\ref{fbl}) by the fermion zero modes 
without taking the asymptotic form.
More conveniently, we 
can use the relation 
\begin{eqnarray}
 \left\langle \frac{\tr \lambda^2}{16 \pi^2} \right\rangle
= \frac{1}{b_0} \Lambda \frac{\partial }{\partial \Lambda}
\langle \frac{1}{2 \pi R} W \rangle,
\end{eqnarray}
which can be derived by uplifting the $\tau$ to 
a superfield
and $\Lambda$ is the dynamical scale in the Pauli-Villars
renormalization scheme at 2-loop order,  
\begin{eqnarray}
 \Lambda^3=\mu^3 \frac{1}{g^2(\mu)} \exp\frac{2 \pi \tau (\mu)}{c_2},
\end{eqnarray}
and $b_0=3 c_2$.
Using this relation,
we finally find
\begin{eqnarray}
 \left\langle \frac{\tr \lambda^2}{16 \pi^2} \right\rangle
=\frac{\Lambda^3 e^{2 \pi i u/c_2}}{\prod_{j=0}^r (k_j^* (\alpha_j)^2/2
)^{k_j^*/2} },
\label{gcc}
\end{eqnarray}
which does not depend on $R$.
For example, 
$\left\langle \frac{\tr \lambda^2}{16 \pi^2} \right\rangle=\pm
\Lambda^3$ 
for $G=SU(2)$.

Now we will consider the original theory at $t=0$.
We will denote $\langle (\cdots ) \rangle_t$ as the correlator of the 
original theory, but choosing 
the vacuum (or the boundary condition at spatial infinity) 
as the one of the theory with the regulator action with $t$.
Note that the vacua of the theory deformed
by the regulator action is not need to be the vacua of the  
original theory in general.
Then, we have 
\begin{eqnarray}
\frac{\partial}{\partial t} 
\langle {\cal O}_1 \cdots {\cal O}_m e^{-t \int \delta V} \rangle_t
=\lim_{\Delta t \rightarrow 0} 
\frac{1}{\Delta t}
\left(
\langle {\cal O}_1 \cdots {\cal O}_m e^{-t \int \delta V}
\rangle_{t+\Delta t}
-\langle {\cal O}_1 \cdots {\cal O}_m e^{-t \int \delta V} \rangle_t
\right)
\end{eqnarray}
where we have used 
$\langle \delta ({\cal O}) e^{-t \int \delta V}  \rangle_t =0 $.
However,
this will diverges if the 
vacuum at $t+\Delta t$ is not the vacuum at $t$.
Now we assume the smoothness of changing the parameter $t$.\footnote{
Although it is expected to be valid from the many examples 
using the localization technique,
this should be justified. 
Unfortunately, for general cases, we can not justify it in this paper.
In our case here we know the correct result, thus this assumption 
will be valid.}
Furthermore, we have seen that the moduli space of vacua of the theory at 
$t \rightarrow \infty$ is discrete.
Thus, the vacuum is independent of $t$, which means that
$\frac{\partial}{\partial t} 
\langle {\cal O}_1 \cdots {\cal O}_m e^{-t \int \delta V} \rangle_t=0$.
Therefore, 
the gaugino condensation (\ref{gcc}),
which is correct value \cite{DHK},
is valid at $t=0$, i.e. the original theory.

\section{Including Chiral Multiplets}

Let us consider the chiral multiplets.
By the regulator action for the vector multiplets,
the effective gauge coupling determined by $t$ 
can be arbitrary weak.
Thus, without introducing the 
regulator action for the chiral multiplets,
the chiral multiplets can be integrated out 
first where the vector multiplets as the background fields.
Indeed, for the (massive) SUSY QCD, 
the matters can be integrated out first trivially and
the dynamical scale of the resulting SUSY Yang-Mills theory 
is computed by the usual way.
Then, we can easily see that 
the resulting gaugino condensation is correct one.

If there is interaction terms in the superpotential, 
the effective superpotential after integrating out the chiral multiplets
is non-trivial function of $S =\frac{1}{32 \pi^2} W_\alpha W^\alpha$.
This is the case for the Dijkgraaf-Vafa conjecture \cite{DV}.
In order to evaluate the gaugino condensation, 
we need to take into account not only the fundamental monopoles,
but general anti-self-dual configurations
because of the interactions which are induced by the chiral multiplets.
This would be rather difficult.
It could be useful to introduce the regulator action
for the chiral multiplets for this case.
We hope to report further progress for this 
in near future.


\section*{Acknowledgments}

S.T. would like to thank 
K. Hosomichi 
for his collaboration at the early stage of
this project and many useful and important comments and discussions. 
S.T. would like to thank  
G. Ishiki and M. Taki also for 
useful discussions.
S.T. would like to thank  E. Poppitz for useful comments
about the 1-loop determinant.
The work of S. T. is partly supported by the Japan Ministry of Education,
Culture, Sports, Science and Technology (MEXT).

\vspace{1cm}


\appendix

\section{4d  ${\cal N}=1$ SUSY Gauge Theory on $S^4$ }

In this appendix, we will explicitly construct 
the SUSY transformations and the SUSY invariant actions for
4d ${\cal N}=1$ SUSY gauge theories on $S^4$.\footnote{
For the chiral multiplets, they 
were already explicitly represented in \cite{Fest}.}
However, as we will see in later,
it may be impossible to construct a SUSY exact regulator term
with a (semi)-positive definite bosonic part
because $(\delta_\xi)^2$ can not be real nor pure imaginary
as shown in \cite{GGK}.


First, 
we will construct ${\cal N}=1$ SUSY theories from 
the ${\cal N}=2$ SUSY theories on $S^4$,
which are realized by a form given in \cite{Te, Nosaka},
by a truncation of the fields.
The notation in this section is the one used in \cite{HST, Te}.
Especially, the indices $\mu,\nu,\cdots$ runs from $1$ to $4$.

The metric of $S^4$ is taken to be
\begin{eqnarray}
ds^2_{S^4} &=& \ell^2(d\theta^2+\sin^2\theta ds^2_{S^3})
~=~ \frac{dr^2+r^2ds^2_{S^3}}{(1+\frac{r^2}{4\ell^2})^2}
~=~ \frac{\sum dx_n^2}{(1+\frac{r^2}{4\ell^2})^2},
\end{eqnarray}
where 
$r^2=\sum_{n=1}^4 (x^n)^2$ and we find 
$e^a=f\delta^a_n dx^n$ with $f \equiv (1+\frac{r^2}{4\ell^2})^{-1}$.
We can embed the $S^4$ in $R^5$ as $Y_1^2+ \cdots + Y_5^2=l^2$.
The relation between $x^n$ and $Y^n$ ($n=1,\ldots,4$)
is $Y_n=\frac{x^n}{1+\frac{r^2}{4 l^2}}$.

We assume the following Killing Spinor equation:
\beq
D_\mu \xi_I = \Gamma_\mu \tilde{\xi}_I.
\eeq
Using the traceless $2 \times 2$ matrix $t_I^J$, which satisfies 
 \begin{eqnarray}
 (t^2)_I^{\,\, J} = \frac{1}{4l^2} \, \delta_I^{\,\,J},
 \end{eqnarray}
the Killing Spinor equation is solved by
\begin{eqnarray}
\xi_I &=& \frac{1}{\sqrt{1+\frac{r^2}{4 l^2}}} \left( 
\epsilon_I+ x^i\Gamma_i \Gamma_5 t_I^{\,\, J} \epsilon_J \right),
\end{eqnarray}
where $i,j=1,\ldots,4$ which are 4d flat indices and 
$\epsilon_I, \epsilon'_I$ are constants, 
and
\begin{eqnarray}
 \tilde{\xi}_I= t_I^{\,\, J} \Gamma_5 \xi_J.
\end{eqnarray}

Therefore, by the $SU(2)_R$ transformation,
we will choose\footnote{
In \cite{Te, HST}, $t \sim \sigma_3$ was chosen.
Our choice here is more convenient for ${\cal N}=1$ SUSY case.
}
\begin{eqnarray}
 t_I^{\,\, J}= \frac{1}{2l} (\sigma_1)_I^{\,\, J}.
\end{eqnarray}

For the ${\cal N}=2$ vector multiplets.
the ${\cal N}=2$ SUSY variation of fields on $S^4$ 
was given by
\begin{eqnarray}
\delta_\xi A_m &=& i\epsilon^{IJ}\xi_I\Gamma_m\lambda_J~~, \nonumber \\
\delta_\xi\sigma &=& i\epsilon^{IJ}\xi_I\lambda_J~~,\nonumber \\
\delta_\xi\lambda_I &=&
-\frac12\Gamma^{mn}\xi_IF_{mn}+\Gamma^m\xi_ID_m\sigma
+\xi_JD_{KI}\epsilon^{JK}+2\tilde\xi_I\sigma~~, \nonumber \\
\delta_\xi D_{IJ} &=&
-i(\xi_I\Gamma^m D_m \lambda_J+\xi_J\Gamma^m D_m \lambda_I)
+[\sigma,\xi_I\lambda_J+\xi_J\lambda_I]
+2 i \xi_K t_{IJ} \Gamma_5 \lambda^K,
\nonumber\\
\label{susytr}
\end{eqnarray}
where $m=1, \ldots,5$, $A_5$ is a scalar, $F_{\mu 5}=D_\mu A_5$ 
and $D_5 (*)=-i [*,A_5]$.
Note that the $SU(2)_R$ transformation including $t_I^J$
acts covariantly on the SUSY transformation.
It was shown that the commutator of the two
SUSY generators is a sum of a translation ($v^m$), a gauge transformation
($\gamma+i v^m A_m$), 
a dilation ($\rho$), an R-rotation ($R_{IJ}$) and 
a Lorentz rotation ($\Theta^{ab}$):
\begin{eqnarray}
~[\delta_\xi,\delta_\eta]A_m &=& -i v^nF_{nm} + D_m\gamma~~, \nonumber \\
~[\delta_\xi,\delta_\eta]\sigma &=& -i v^nD_n\sigma 
+\rho\sigma~~, \nonumber \\
~[\delta_\xi,\delta_\eta]\lambda_I &=&
-i v^n \nabla_n \lambda_I+i[\gamma,\lambda_I]+\frac32\rho\lambda_I
+{R'}_I^{~J}\lambda_J
+\frac14\Theta^{ab}\Gamma^{ab}\lambda \nonumber \\
 &=&
-i v^n D_n \lambda_I+i[\gamma,\lambda_I]+\frac32\rho\lambda_I
+{R}_I^{~J}\lambda_J
+\frac14\Theta^{ab}\Gamma^{ab}\lambda~~, \nonumber \\
~[\delta_\xi,\delta_\eta]D_{IJ} &=&
-i v^n \nabla_n D_{IJ}+i[\gamma,D_{IJ}]+2\rho D_{IJ}
+{R'}_I^{~K}D_{KJ}+{R'}_J^{~K}D_{IK} \nonumber \\
 &=&
-i v^n D_n D_{IJ}+i[\gamma,D_{IJ}]+2\rho D_{IJ}
+{R}_I^{~K}D_{KJ}+R_J^{~K}D_{IK}
~~,
\label{com}
\end{eqnarray}
where $R_I^{~J}=\epsilon^{JK}R_{IK}$ and
\begin{eqnarray}
v^m &=& 2\epsilon^{IJ}\xi_I\Gamma^m\eta_J~~, \nonumber \\
\gamma &=& -2i\epsilon^{IJ}\xi_I\eta_J\sigma~~, \nonumber \\
\rho &=& -2i\epsilon^{IJ}(\xi_I\tilde\eta_J-\eta_I\tilde\xi_J)=0~~,
\nonumber \\
R_{IJ} &=&   4 i t_{IJ} \epsilon^{KL} 
             \xi_K \Gamma_5 \eta_L ~~, \nonumber \\
\Theta^{ab} &=&
-2i\epsilon^{IJ}(\tilde\xi_I\Gamma^{ab}\eta_J-\tilde\eta_I\Gamma^{ab}\xi_J)~~.
\label{intsym}
\end{eqnarray}

Now let us consider the hypermultiplets,
The system of
$r$ hypermultiplets consists of scalars $q^A_I$, fermions $\psi^A$ and
auxiliary scalars $F^A_I$. Here, $I=1,2$ is the $SU(2)$ R-symmetry index
and $A=1,\cdots,2r$. The fields obey the reality conditions
\begin{equation}
(q^A_I)^\ast ~=~ \Omega_{AB}\epsilon^{IJ}q^B_J~~,\quad
(\psi^{A\alpha})^\ast~=~ \Omega_{AB}C_{\alpha\beta}\psi^{B\beta}~~,\quad
(F^A_I)^\ast ~=~ \Omega_{AB}\epsilon^{IJ}F^B_J~~,
\end{equation}
where $\epsilon^{IJ},C_{\alpha\beta}, \Omega_{AB}$ are antisymmetric
invariant tensors of $SU(2)\simeq Sp(1), Spin(5)\simeq Sp(2)$ and
the ``flavor symmetry'' of $r$ free hypermultiplets $Sp(r)$. The
coupling to vector multiplets can be introduced via gauging a subgroup of
$Sp(r)$.
In the Euclidian signature, 
we regard the fields are holomorphic variables, and then
we will forget these reality conditions.
(Note that two complex fields with a reality condition 
have two real components and two holomorphic fields without conditions
also have two components.)
To introduce the coupling to gauge fields and other fields in
the vector multiplet, we need first to introduce the covariant derivative
\begin{equation}
D_m\psi^A ~\equiv~ \partial_m\psi^A-i(A_m)^A_{~B}\psi^B,~\text{etc.}
\end{equation}
Requiring $\Omega_{AB}$ to be gauge-invariant, one finds
$(A_m)_{AB}\equiv\Omega_{AC}(A_m)^C_{~B}$ to be symmetric in the indices $A,B$.

The ${\cal N}=2$ SUSY transformation was given by
\begin{eqnarray}\label{hypert}
\delta q_I &=& -2i\xi_I\psi,\nonumber \\
\delta\psi &=&
\epsilon^{IJ}\Gamma^m\xi_ID_mq_J
+i\epsilon^{IJ}\xi_I\sigma q_J
+2 \epsilon^{IJ} \tilde{\xi}_I q_J
+\epsilon^{I' J'}\check\xi_{I'}F_{J'},\nonumber \\
\delta F_{I'} &=&
2\check\xi_{I'}(i\Gamma^mD_m\psi+\sigma\psi+\epsilon^{KL}\lambda_Kq_L).
\end{eqnarray}
Here, $\check\xi_{I'}$ is a constant spinor which satisfies
\begin{equation}
\epsilon^{IJ}\xi_I\xi_J=\epsilon^{I' J'}\check\xi_{I'}\check\xi_{J'}~~,\quad
\xi_I\check\xi_{J'}=0~~,\quad
\epsilon^{IJ}\xi_I\Gamma^m\xi_J+\epsilon^{I' J'}\check\xi_{I'}\Gamma^m\check\xi_{J'}=0~~.
\end{equation}
The square of $\delta$ is
\begin{eqnarray}
\delta^2 q_I &=& i v^m D_mq_I -i \gamma q_I
-R_I^{\,\,\, J} q_J
\nonumber \\
\delta^2\psi &=& i v^m D_m\psi-i \gamma\psi
-\frac{1}{4} \Theta^{ab} \Gamma^{ab} \psi
\nonumber \\
\delta^2 F_{I'} &=& i v^m D_m F_{I'}-i \gamma F_{I'}
+{R'}_{I'}^{\,\,\, J'} F_{J'}~~,
\end{eqnarray}
where 
\begin{eqnarray}
v^m &=& \epsilon^{IJ}\xi_I\Gamma^m\xi_J~~, \nonumber \\
\gamma &=& -i\epsilon^{IJ}\xi_I\xi_J\sigma~~, \nonumber \\
R_{IJ} &=& 
2 i (\epsilon^{K L} \xi_K \Gamma^5 t_{IJ} \xi_L) 
~~, \nonumber \\
\Theta^{ab} &=&
-2i\epsilon^{IJ} \tilde\xi_I\Gamma^{ab}\xi_J
~~, \nonumber \\
R'_{I' J'}&=& -2i\check\xi_{I'}\Gamma^mD_m\check\xi_{J'}~~,
\end{eqnarray}
which is consistent with the one for the vectormultiplets.


Now we take a SUSY generator with a Killing spinor 
which satisfies\footnote{
We can not impose the twisted Majorana condition for these.}
\begin{eqnarray}
 \Gamma_5 \epsilon_1 = \epsilon_1, \,\,\, \Gamma_5 \epsilon_2=-\epsilon_2,
\end{eqnarray}
which is equivalent to 
\begin{eqnarray}
 P \epsilon_I = \epsilon_I,
\end{eqnarray}
where $P \equiv \Gamma_5 (\sigma_3)_I^{\,J}$.
Then, we find 
\begin{eqnarray}
 \Gamma_5 \xi_1 = \xi_1, \,\,\, \Gamma_5 \xi_2=-\xi_2,
\end{eqnarray}
i.e. $P \xi_I =\xi_I$.

Because $\eta_- \xi_+=0$ for arbitrary chirality $+$ and $-$ spinors,
these Killing spinors satisfy the followings:
\begin{eqnarray}
 \epsilon^{IJ} \xi_I \xi_J=0, \,\,\,\,
v^5=2 \xi_1\Gamma^5 \xi_2=-2 \xi_1 \xi_2=0, \,\,\,\,
R_{IJ}=-4 t_{IJ} \xi_1 \xi_2=0.
\end{eqnarray}
This implies that we can take 
\begin{eqnarray}
 \check\xi_{I'} = i \xi_I.
\end{eqnarray}
For this choice, $R'_{{I'} {J'}}=0$.

For the scalars and vectors,
we will define the action of $P$
as 
\begin{eqnarray}
 P A_\mu =A_\mu, \,\,\, P D_{12} =D_{12}, \,\,\,\, \CR
P A_5 =-A_5, \,\, P \sigma =- \sigma, \,\,
P D_{11} =-D_{11}, \,\, P D_{22} =-D_{22}.
\end{eqnarray}
Then, we find that 
\begin{eqnarray}
 [\delta_\xi, P]=0.
\end{eqnarray}
This means that, schematically,
$\delta_\xi \Phi_+ = \Phi_+ + \Phi_+ \Phi_+ + \Phi_- \Phi_- $
and 
$\delta_\xi \Phi_- = \Phi_- + \Phi_- \Phi_+ $.

Now, we consider only the $P=1$ fields as an ${\cal N}=1$ SUSY fields.
The SUSY transformation on the ${\cal N}=1$ SUSY fields are defined
just by taking $P=-1$ fields vanish in the SUSY transformation of ${\cal N}=2$ SUSY.
Thus, the SUSY algebra is obtained by 
taking $P=-1$ fields vanish in the one for ${\cal N}=2$ SUSY,
which is consistent.

Explicitly, we find
\begin{eqnarray}
\delta_\xi A_\mu &=& i (\xi_1 \Gamma_\mu \lambda_2^- ~~ 
-\xi_2 \Gamma_\mu \lambda_1^+) ~~, \nonumber \\
\delta_\xi \lambda_1^+ &=&
-\frac12\Gamma^{\mu \nu}\xi_1 F_{\mu \nu}+\xi_1 D_{12}~~, \nonumber \\
\delta_\xi \lambda_2^- &=&
-\frac12\Gamma^{\mu \nu}\xi_2 F_{\mu \nu}-\xi_2 D_{12}~~, \nonumber \\
\delta_\xi D_{12} &=&
-i \xi_1 \Gamma^\mu D_\mu \lambda_2^-
-i \xi_2 \Gamma^\mu D_\mu \lambda_1^+ ,
\nonumber\\
\label{susytr1}
\end{eqnarray}
where 
$\lambda_1^+= (1+\Gamma_5)/2 \,\, \lambda_1$ 
and $\lambda_2^-= (1-\Gamma_5)/2 \,\, \lambda_2$.

If we define $D \equiv D_{12}$, $\xi \equiv \xi_1+\xi_2$
and $\lambda \equiv \lambda_1^+ +\lambda_2^-$,
we find 
\begin{eqnarray}
\delta_\xi A_\mu &=& -i \xi \Gamma_\mu  \Gamma_5 \lambda, \nonumber \\
\delta_\xi \lambda &=&
-\frac12 \Gamma^{\mu \nu}  \xi F_{\mu \nu}+\Gamma_5 \xi D~~, \nonumber \\
\delta_\xi D &=&
-i \xi \Gamma^\mu D_\mu \lambda,
\label{susytr1a}
\end{eqnarray}
which is same as the one in flat space
except that
the vierbein and the connections are the ones on $S^4$.

For the hypermultiplets,
we can take 
\begin{eqnarray}
\Gamma_5 \check\xi_1 = \check\xi_1, \,\, \Gamma_5 \check\xi_2 = -\check\xi_2.
\end{eqnarray}
We will take 
\begin{eqnarray}
\Omega_{AB}=i \sigma_2 \oplus i \sigma_2 \oplus \cdots \oplus i \sigma_2,
\end{eqnarray}
and we will change the notation for the index $A$, for example,
\begin{eqnarray}
 \psi^A \rightarrow \psi^{ A, i },
\end{eqnarray}
now the $A=1,\ldots,r$ and $i=1,2$.
Then, we define 
$P \psi^{A, i} \equiv \Gamma_5 (-1)^{i+1} \psi^{A i}$,
$P q_I^{A,i}=(-1)^{I+i} q_I^{A,i}$
and 
$P F_I^{A,i} = -(-1)^{I+i} F_I^{A+i}$.
With this, we can consistently truncate the ${\cal N}=2$ fields to 
the $P=1$ fields, i.e.  ${\cal N}=1$ fields,
and find
\begin{eqnarray}\label{hypert1}
\delta q^{A,i} &=& -2i\xi_i \psi^{A,i},\nonumber \\
\delta \psi^{A,i} &=&
- \Gamma^\mu \epsilon^{ij} \xi_j D_\mu q^{A,i}
- \frac{1}{l}  \xi_i q^{A,i}
+\check\xi_{i}F^{A,i},\nonumber \\
\delta F^{A,i} &=&
2 \epsilon^{ij} \check\xi_{j} (i\Gamma^\mu D_\mu \psi^{A,i} 
-\epsilon^{ik} \lambda_k q^{A,i}
),
\end{eqnarray}
where we have not summed over $i$ and 
we defined $q^{A,i} \equiv q^{A,i}_i$ (no summation for $i$), 
$F^{A,i} \equiv (-1)^{i-1} F^{A,i}_{3-i}$
and $\psi^{A,i} \equiv (1 +(-1)^{i+1} \Gamma_5)/2 \, \psi^{A,i}$.
This form is slightly different from the one in \cite{Fest}.
For $\check\xi_{I'}=i \xi_I$, however, 
if we define 
\begin{eqnarray}
F'=F+\frac{i}{l} q, 
\end{eqnarray}
we obtain the same form.

We will regard the fields as holomorphic and then 
forget the reality conditions. Indeed, we will not encounter
any complex conjugate of the fields below.


Below
we will try to construct 
SUSY invariant actions.
We will drop the total divergent terms below for the notational
convenience.

The ${\cal N}=2$ SUSY invariant action for vectormultiplets on $S^4$
is
\begin{eqnarray}
{\cal L}_{S^4}^{vector} &=& \frac12 F_{mn} F^{mn} - D_m \sigma D^m \sigma
+ i \lambda_I \Gamma^m D_m \lambda^I-\lambda_I [\sigma,\lambda^I] \CR
&&
-\frac12 (D_{IJ}-2 A_5 t_{IJ} )(D^{IJ}-2 A_5 t^{IJ})
-4 t_{IJ} t^{IJ} \left( (A_5)^2-\sigma^2) \right), \CR
\end{eqnarray}
where $\partial_5=0$, which is the usual SUSY Yang-Mills 
Lagrangian of the vector multiplet
used in \cite{Pestun} with some field redefinitions \cite{Te}.
This action does not have any terms linear in fields with $P=-1$.
Thus, the following truncated action is invariant under the ${\cal N}=1$ SUSY:
\begin{eqnarray}
{\cal L}_{S^4}^{vector} &=& \frac12 F_{\mu \nu} F^{\mu \nu} 
+ i \lambda_1^+ \Gamma^\mu D_\mu \lambda_2^-  
- i \lambda_2^- \Gamma^\mu D_\mu \lambda_1^+  
-(D_{12} )^2 , 
\end{eqnarray}
which take the same form in the flat space.

Now we will consider the hypermultiplets.
We have the ${\cal N}=2$ SUSY invariant Lagrangian on $S^4$:
\begin{eqnarray}
\label{hyperla4d}
{\cal L}^{hyper}_{S^4}
&=& \epsilon^{IJ}(D_\mu \bar q_I D^\mu q_J
+\bar q_I (A_5)^2 q_J
-\bar q_I\sigma^2 q_J
)
-2(i \bar{\psi} \Gamma^\mu D_\mu \psi
+\bar\psi \Gamma_5 A_5 \psi
+\bar\psi\sigma\psi
)
\nonumber \\ &&
-i\bar q_I {D'}^{IJ}q_J-4\epsilon^{IJ}\bar\psi\lambda_Iq_J
-\epsilon^{I' J'}\bar F_{I'}F_{J'} \CR
&& - 8 t^{KL}t_{KL}\epsilon^{IJ}\bar q_Iq_J
~~,
\end{eqnarray}
where, we have introduced the notation $\bar\psi_B\equiv\psi^A\Omega_{AB}$
and suppress the indices $A,B,\cdots$, such that
\begin{eqnarray}
\epsilon^{IJ}\Omega_{AB}D_\mu q^A_I D^\mu q^B_J &\equiv&
\epsilon^{IJ}D_\mu \bar q_I D^\mu  q_J~~,\nonumber \\
\Omega_{AB}\psi^A\Gamma^\mu (A_\mu )^B_{~C}\psi^C &\equiv&
\bar\psi\Gamma^\mu A_\mu \psi~~,
~~\text{etc.}
\end{eqnarray}

Because this action does not have any term which is linear in
the $P=-1$ fields,
the following action is ${\cal N}=1$ SUSY invariant:
\begin{eqnarray}
\label{chirala1}
{\cal L}^{chiral}_{S^4}
&=& 2 h_{AB} D_\mu q^A D^\mu q^B
-2i \Omega_{AB} \bar{\psi}^A \Gamma^\mu D_\mu \psi^B \CR
&& 
-2 i h_{AB} q^A {D}^{12} q^B -4 h_{AB} \psi^A (\lambda_{A \, mod \, 2}
\, q)^B
-2 h_{AB} F^A F^B 
- 16 t^{KL}t_{KL} \, h_{AB} q^A q^B
~~, \nonumber \\
\end{eqnarray}
where 
\begin{eqnarray}
 h_{AB} \equiv \Omega_{AB} (-1)^{A}
=\sigma_1 \oplus \sigma_1 \oplus \cdots  \oplus \sigma_1.
\end{eqnarray}

We can easily see that 
the flat space superpotential terms written in 
$F'=F+\frac{i}{l} q$ is invariant under 
the ${\cal N}=2$ SUSY on $S^4$
because the superpotential is gauge invariant and
the $F'$ enters in the superpotential at most linearly.
Note that the SUSY transformation is modified 
only for $F'$ except the modification of the metric and connections.
The SUSY transformation $\check\xi_2 \Gamma_\mu D_\mu \psi$ in $F'$
gives an extra contribution $\- D_\mu \check\xi_2 \Gamma_\mu \psi$,
which indeed cancel with the one from the modified term in $F'$.
The theta term is also ${\cal N}=1$ SUSY invariant because it is a topological term.


Let us consider SUSY invariant operators.
Because $\xi_1=0$ at $x^\mu=0$, we find
$\delta_\xi \lambda_1^+ (x^\mu=0) =0$,
$\delta_\xi q^{A,1} (x^\mu=0) =0$,
and $\delta_\xi F^{A,2} (x^\mu=0) =0$.
Similarly we find $\delta_\xi \lambda_2^- (x^\mu=\infty) =0$,
$\delta_\xi q^{A,2} (x^\mu=\infty) =0$,
and $\delta_\xi F^{A,1} (x^\mu=\infty) =0$.

Note that $q^{A,2}$ can be considered as $(q^{A,1})^\dagger$,
thus the lowest components of the chiral superfields
inserted at the north pole and 
the lowest components of the anti-chiral superfields
inserted at the south pole are the SUSY invariants operators.
(In the flat case, 
it is clear that $\bar{D}^2 (\Phi^\dagger)\sim \bar{F}$ is a chiral operator
because $\bar{D}^3=0$.)


Now, we will try to apply
the localization technique used in \cite{Pestun}
to the ${\cal N}=1$ SUSY theory on $S^4$.
However, as we will see below, it may be impossible
to construct a term $\int_{S^4} \delta V$ which 
has a positive definite real part.

First we will try to construct $\int_{S^4} \delta V$ on $S^4$
with a appropriate properties.
We take $\xi_I$ as Grassmann-even spinors such that $\delta_\xi$
is a fermionic transformation.
We can easily see that $\xi_1 \xi_1=0$,
which is followed form $C^T=-C$ and $\xi_1 \xi_2=0$ because 
$\Gamma_5 \xi_1=\xi_1$ and $\Gamma_5 \xi_2=-\xi_2$.
In order to use the localization technique,
we need a regulator Lagrangian $\delta_\xi V$ with
$\int_{S^4} (\delta_\xi)^2 V=0$.
The usual choice is 
a form like $V = \text{tr}\big[(\delta_\xi \lambda)^\dagger
\lambda\big]$,
where $(\delta_\xi \lambda)^\dagger$ should be defined 
using the holomorphic fields and $\xi_I$.
However, if we assume the following form
\begin{eqnarray}
 (\xi_I)^\star =(M^{IJ}+ N^{IJ} \Gamma_5) C \xi_J,
\end{eqnarray}
where $M,N$ are arbitrary matrices,
we find a contradiction.
Indeed, 
the $\Gamma_5$-chirality requires that $M,N$ are diagonal,
and we can assume $ (\xi_I)^\star =M C \xi_I$.
Then, we find
$\xi_I=M^* C^* (\xi_I)^*
=M^* C^* M C \xi_I
=-|M|^2 \xi_I$, which means $\xi_I=0$.
We can use a tensor satisfying ${\cal L}_v T=0$, however,
as far as we have checked, there is 
no localization terms with a positive definite real part of bosonic terms.
These difficulty will be originate from the nonzero complex value 
of the $v^{\mu}= \xi_I \Gamma^{\mu} \xi^I$
even if we take $\epsilon_1=0$.

\end{document}